\documentclass[journal=nalefd,manuscript=article]{achemso}
\usepackage[normalem]{ulem}
\usepackage[utf8]{inputenc} 
\usepackage{xcolor}
\usepackage[version=3]{mhchem} 
\usepackage[T1]{fontenc}       

\author{Matthias B. Jungfleisch}
\affiliation[]{Materials Science Division, Argonne National Laboratory, Argonne, Illinois 60439, USA}
\email{jungfleisch@anl.gov}
\author{Junjia Ding}
\affiliation[]{Materials Science Division, Argonne National Laboratory, Argonne, Illinois 60439, USA}
\author{Wei Zhang}
\affiliation[]{Materials Science Division, Argonne National Laboratory, Argonne, Illinois 60439, USA}
\alsoaffiliation[]{{Department of Physics, Oakland University, Rochester MI , 48309, USA}}
\author{Wanjun Jiang}
\affiliation[]{Materials Science Division, Argonne National Laboratory, Argonne, Illinois 60439, USA}
\alsoaffiliation[]{State Key Laboratory of Low-Dimensional Quantum Physics and Department of Physics, Tsinghua University, and Collaborative Innovation Center of Quantum Matter, Beijing 100084, China}
\author{John E. Pearson}
\affiliation[]{Materials Science Division, Argonne National Laboratory, Argonne, Illinois 60439, USA}
\author{Valentine Novosad}
\affiliation[]{Materials Science Division, Argonne National Laboratory, Argonne, Illinois 60439, USA}
\author{Axel Hoffmann}
\affiliation[]{Materials Science Division, Argonne National Laboratory, Argonne, Illinois 60439, USA}

\title{Insulating nanomagnets driven by spin torque}


\begin{document}

%
%
%
%
%

\begin{abstract}
Magnetic insulators, such as yttrium iron garnet (Y$_3$Fe$_5$O$_{12}$), are ideal materials for ultra-low power spintronics applications due to their low energy dissipation and efficient spin current generation and transmission. Recently, it has been realized that spin dynamics can be driven very effectively in micrometer-sized Y$_3$Fe$_5$O$_{12}$/Pt heterostructures by spin-Hall effects. We demonstrate here the excitation and detection of spin dynamics in Y$_3$Fe$_5$O$_{12}$/Pt nanowires by spin-torque ferromagnetic resonance. The nanowires defined via electron-beam lithography are fabricated by conventional room temperature sputtering deposition on Gd$_3$Ga$_5$O$_{12 }$ substrates and lift-off. We observe field-like and anti-damping-like torques acting on the magnetization precession, which are due to simultaneous excitation by Oersted fields and spin-Hall torques. The Y$_3$Fe$_5$O$_{12}$/Pt nanowires are thoroughly examined over a wide frequency and power range. We observe a large change in the resonance field at high microwave powers, which is attributed to a decreasing effective magnetization due to microwave absorption. {These heating effects are much more pronounced in the investigated nanostructures than in comparable micron-sized samples.} By comparing different nanowire widths, the importance of geometrical confinements for magnetization dynamics becomes evident{: quantized spin-wave modes across the width of the wires are observed in the spectra.} Our results are the first stepping stones toward the realization of 
integrated magnonic logic devices based on insulators, where nanomagnets play an essential role.

\end{abstract}

\section{Introduction}

Magnon spintronics is a subfield of spintronics concerned with the exploration of magnons, the elementary quanta of spin waves, as carriers of spin-angular momentum. \cite{Chumak_NatPhys_2015} This generated renewed interest in the ferrimagnetic insulator yttrium iron garnet (YIG, Y$_3$Fe$_5$O$_{12}$), due to it's extremely low Gilbert damping. {Due to its small damping, it is easily possible to create spin-wave spin currents and transmit them over relatively large distances of up to cm length scales in thick YIG. Ultimately, the utilization of those pure spin currents which are not accompanied by electronic charge currents might open the way to low-power spintronics based on magnetic insulators.} Because of its low dissipation, YIG has been used for many decades as a key component in microwave and radar technologies. In the past, YIG single crystal were mainly grown by liquid phase epitaxy (LPE), which yields films in the thickness range from several hundreds of nanometers to millimeters. Motivated by recent discoveries of magnon spintronics effects in YIG-based heterostructures such as spin-tranfer torque effects, \cite{Jungfleisch_PRL_2016,Sklenar_PRB_2015,Lauer_APL_2016,Schreier_PRB_2015,Kajiwara_Nature_2010,Collet_Nat_Com_2016,Hamadeh_PRL_2014} magnetic logics \cite{Chumak_NatPhys_2015} and auto-oscillations\cite{Hamadeh_PRL_2014,Collet_Nat_Com_2016}, there is an urgent need for miniaturization {which is indispensable to integrate YIG in real spintronics application. This} calls for alternative approaches to grow YIG films in the nanometer range. It was shown that high-quality nanometer-thick YIG films can be grown by LPE \cite{Castel_PRB_2012}, pulsed laser deposition \cite{Jungfleisch_PRB_2015,Onbaslli_APL_Mat_2014,Lin_APL_2013} and sputtering deposition \cite{Liu_JAP_2014,Chang_IEEE_2014,Sun_APL_2012} and even the micro- and nanostructuring of YIG samples was reported. \cite{Sklenar_PRB_2015,Jungfleisch_PRL_2016,Li_Nanoscale_2016,Jungfleisch_JAP_2015}

The most common way to excite and detect magnetization dynamics is by inductive techniques using conventional antenna structures. Besides that, the combination of spin pumping and inverse spin Hall effect has become a standard method to detect spin dynamics in magnetic insulators. The generation of spin dynamics by the reciprocal effects, however, is more complicated. \cite{Wang_APL_2011,Hahn_PRB_2013} Reasons for that are the interface nature of the involved spin-transfer torque process that makes it difficult to drive thick insulating magnets. Furthermore, it is important that the damping of the magnetic material is low enough so that the threshold for the onset of auto-oscillations can be surpassed easily \cite{Hamadeh_PRL_2014} and that the mode spectrum is discrete enabling a selective pumping of spin-wave modes. \cite{Jungfleisch_PRB_2015} Another way to drive and detect magnetization dynamics by electrical means is spin-torque ferromagnetic resonance (ST-FMR), a method that was originally developed for all-metallic systems. \cite{Liu_PRL_2011} Recently, it was shown that this method can also be used in magnetic insulators, \cite{Jungfleisch_PRL_2016,Sklenar_PRB_2015,Chiba_PRA_2014,Chiba_JAP_2015,Schreier_PRB_2015} {where $dc$ voltage detection of magnetization dynamics is achieved by a simultaneous rectification mechanism due to spin pumping/inverse spin Hall effect and a mixing of spin-Hall magnetoresistance (SMR) with the microwave signal. SMR is a magnetoresistance effect that stems from the magnetization-direction dependent spin current backflow at the interface between a magnet and a normal metal and is based on the spin-Hall effect} \cite{Nakayama_PRL_2013}. Full angular-dependent studies of ST-FMR in YIG/Pt microstructures were performed. Furthermore, investigations by Brillouin light scattering spectroscopy \cite{Jungfleisch_PRL_2016} revealed the mode profile of the ST-FMR driven spin-wave mode with nonlinear effects. However, in order to make use of these intriguing spin-wave phenomena in real high-frequency applications further miniaturization to the sub-micron level is urgently required.

\begin{figure}
  \includegraphics[width=0.65\columnwidth]{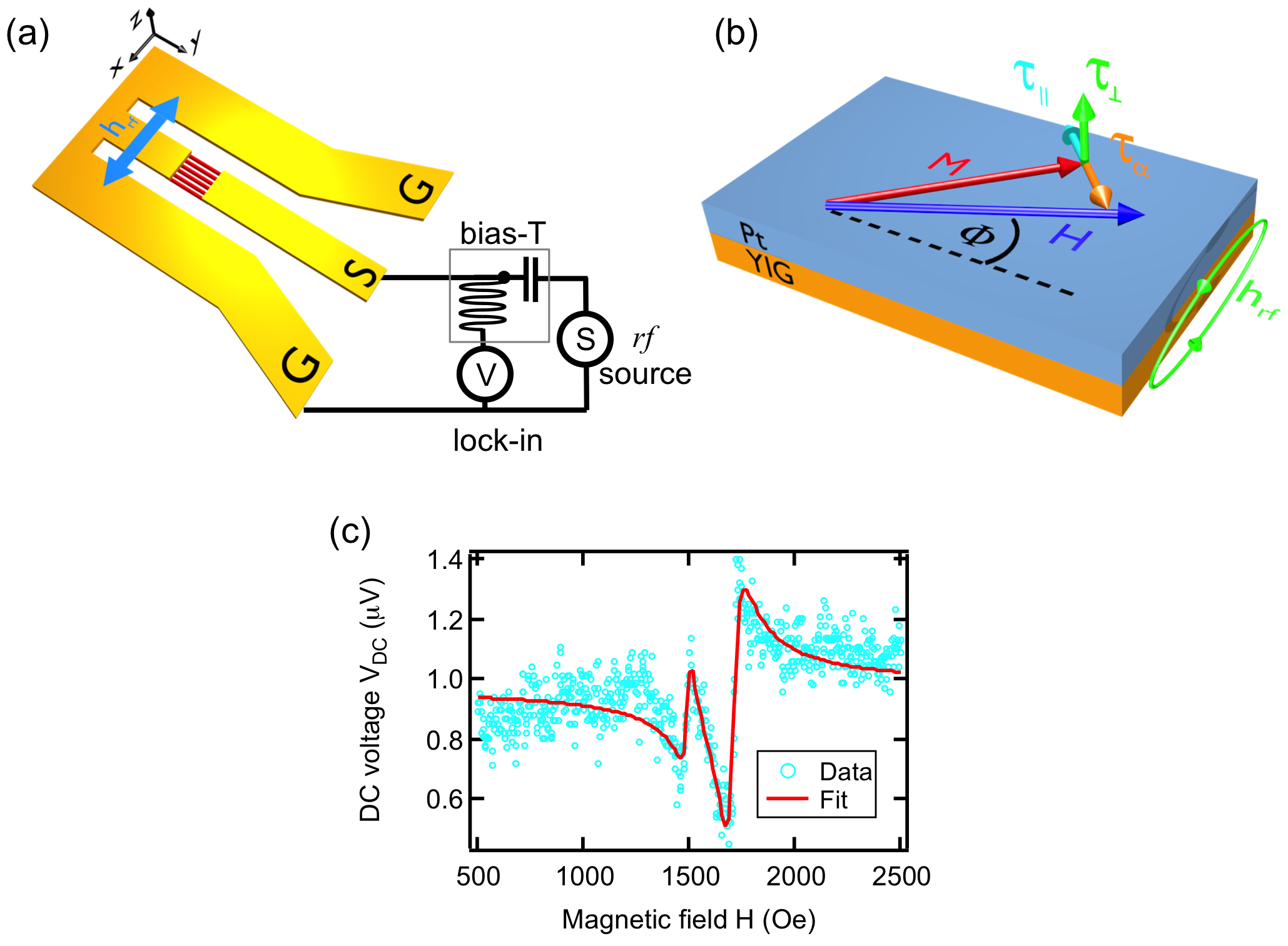}
  \caption{(a) Schematic of the ST-FMR experimental setup. A bias-T is utilized for simultaneous $rf$ transmission and $dc$-voltage detection by lock-in techniques. (b) Illustration of ST-FMR mechanism: The alternating charge current in the Pt layer drives the magnetization to resonance by a field-like ($\tau_\perp$) and an anti-damping like ($\tau_\parallel$) torque. The external magnetic field is applied at an angle $\phi = 45^\circ$. (c) Typical ST-FMR spectrum. The figure shows the spectrum of a wire width of 365 nm at 7 GHz and 1 mW.}
  \label{Fig_Setup}
\end{figure}

In this Letter, we show the realization of nanoscaled spin-torque ferromagnetic resonance devices based on YIG. The epitaxial patterned YIG nanowires are fabricated using electron beam lithography, lift-off and post annealing, followed by another electron beam lithography step to define the Pt strips. \cite{Li_Nanoscale_2016} We observe superimposed symmetric and antisymmetric Lorentzian lineshapes due to simultaneous excitation of spin dynamics by Oersted field and spin-Hall spin-transfer torque effect. The rectified $dc$ voltage spectra are studied over a wide frequency and power range and we perform microwave-power dependent measurements that indicate the occurrence of strong heating effects {in nanometer size YIG/Pt samples}. By comparing different nanowire widths, the importance of geometrical confinements for magnetization dynamics becomes evident{, which could not be observed in micrometer size samples \cite{Jungfleisch_PRL_2016}.} Besides that low temperature measurements give further insights into the spin dynamics in magnetic insulators driven by ST-FMR.

\section{Device Fabrication}

\begin{figure}
  \includegraphics[width=1\columnwidth]{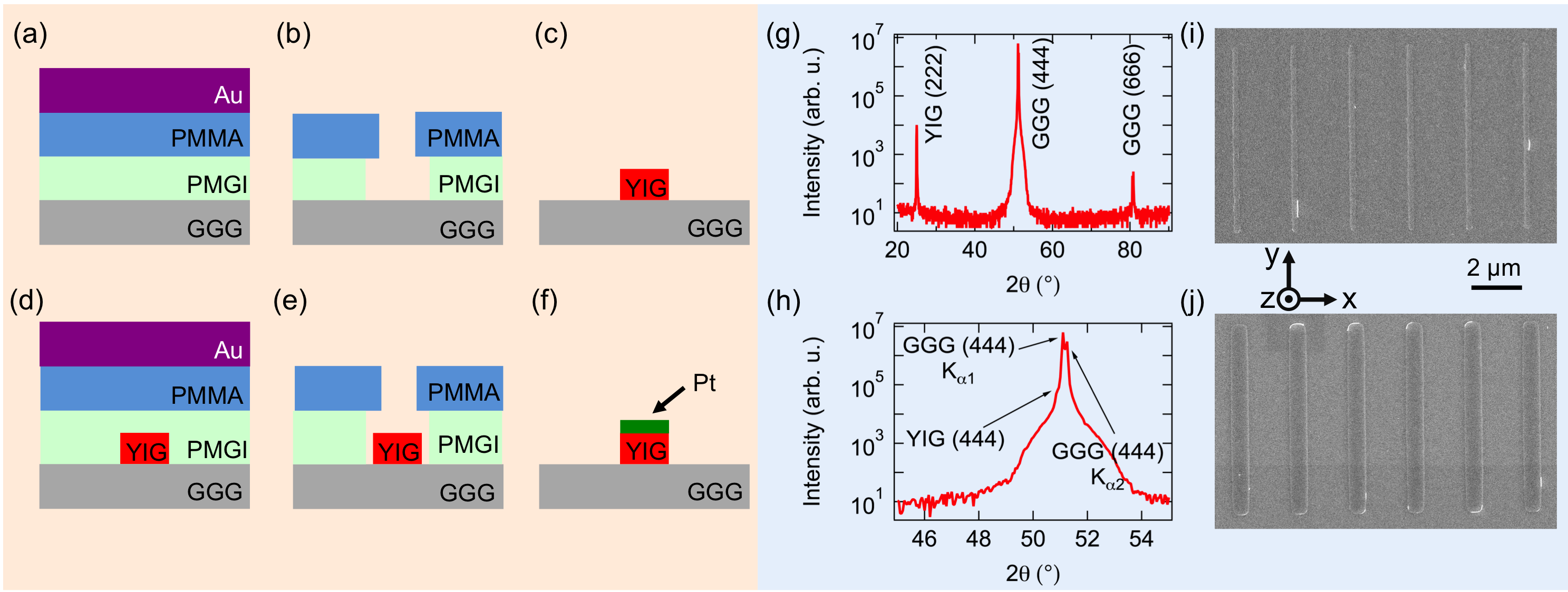}
  \caption{Illustration of the multi-step electron beam lithography process. (a) Since the GGG substrate is insulating a 5-nm-thick Au layer is deposited on the PMMA/PMGI bilayer resist. (b) Removing Au by a gold etcher and development. The utilization of a bilayer resist leads to a undercut. (c) After YIG deposition and lift-off, the samples are annealed $ex-situ$. (d) The same bilayer resist is used for the second electron beam lithography step. (e) Removing Au and development. (e) Pt deposition and lift-off. The sample is now ready for a final optical lithography process to define the shortened coplanar waveguide that serves as a electrical lead. (g,h) X-ray diffraction pattern of a 30-nm thick annealed YIG film and the same spectrum in a magnified scale showing the YIG (444) peak, (h). (i,j) Scanning electron microscopy images of patterned YIG nanowires on GGG substrates before Pt strips are patterned on the top. (i) 365-nm wide wires, (j) 765-nm wide wires.}
  \label{Fig1}
\end{figure}

We use a multistep lithography process to fabricate the YIG/Pt ST-FMR nanodevices \cite{Li_Nanoscale_2016}. YIG (thickness 30 nm) is deposited on insulating (111)-oriented gadolinium gallium garnet (Gd$_3$Ga$_5$O$_{12}$, GGG) single crystal substrates at room temperature (RT) from a stoichiometric YIG target. Optimal growth condition at RT is achieved for an Ar gas flow of 16 sccm, a base pressure of 10 mTorr and a sputtering power of 75 W, which results in a low sputtering rate of \textless 0.25 \AA/s. Prior to the sputtering process, we define the nanostructures onto PMMA/PMGI bilayer resist via electron beam lithography (Raith eLINE Plus). The utilization of PMMA/PMGI bilayer resists facilitates the formation of an undercut crosssection profile which is beneficial for magnetron sputtering deposition \cite{Zhang_2014}. Since the GGG substrates are highly insulating, a 5-nm-thick Au layer is deposited onto the bilayer resist before electron beam exposure. The thin Au layer allows for an effective electron charge dissipation during the lithography process. In a subsequent step, the Au layer is removed after exposure by using a gold etcher. The resist is developed using MicroChem MIBK (PMMA) and Shipley CD-26 (PMGI), respectively. After deposition of 30 nm YIG, the resist is lifted-off by Shipley 1165 and all surplus material is removed except the nanonwire structures. The samples are subsequently annealed \textit{ex-situ} at 800$^\circ$C for 2 hours in a tube furnace with continuous air flow. The temperature is ramped up at a rate of 180$^\circ$C/h and ramped down at a rate of 120$^\circ$C/h. Figure~\ref{Fig1}(i) and (j) show the morphology of the nanostructured YIG  by using scanning electron microscopy after 5 nm Au deposition. We investigate two sets of nanowires: the first set consists of 6 parallel wires having a width of 365~nm [Fig.~\ref{Fig1}(i)], the second set consists of 6 parallel wires with a width of 765~nm [Fig.~\ref{Fig1}(j)]. After the YIG nanowire fabrication, PMMA/PMGI bilayer resist is spin coated and a 5-nm Au layer is deposited onto the samples, followed by another electron beam lithography process, in which the Pt bars are defined on top of the YIG nanowires. A precise alignment is required in this step. After removing the Au and developing the resist as described above, we use plasma cleaning at 20 W for 120 s to clean the YIG surface \cite{Jungfleisch_APL_2013}, followed by deposition of 5 nm Pt using magnetron sputtering and lift-off. {The measurements were performed on the collection of those wires. Calculations confirm that the separation} between the wires is large enough to avoid any crosstalk between the wires. {The width variation among the different wires of one set is less than 2\% and the variation of one individual wire along its length is less than 5\%. Thus, the dynamics field distribution is not affected.}  Finally, electrical leads made of Ti(3 nm)/Au(120 nm) are defined via optical lithography{, see Fig.~\ref{Fig_Setup}(a). The width of the center arm is 16 $\mu$m, the width of the outer arms is 35 $\mu$m. They are separated by a 42 $\mu$m gap. The shortened end of the waveguide has a width of 25 $\mu$m. The ends of center arm cover approximately 1 $\mu$m of the nanowires to ensure good electrical contact.} 

For comparison, we characterize our reference YIG film that was not nanostructured. Fig.~\ref{Fig1}(g) and (h) illustrate X-ray diffraction patterns of our bare YIG films. The data confirm epitaxial (111)-oriented YIG growth on the GGG substrates. Vibrating sample magnetometry with an in-plane magnetic field reveals a very small coercivity below 5 Oe. The characterization of the dynamic properties can be found in Ref.~\cite{Li_Nanoscale_2016}.

 \section{Experimental Setup}
 
A standard ST-FMR configuration is used as  illustrated in Fig.~\ref{Fig_Setup}. We measure the spectra using a bias-T allowing for simultaneous transmission of microwave signals with $dc$ voltage detection via lock-in technique across the Pt. The amplitude of the $rf$ current is modulated at 2 kHz. The external magnetic field is applied at an angle of $\phi \approx 45^\circ$ where the $dc$ voltage detection is maximized. \cite{Jungfleisch_PRL_2016,Sklenar_PRB_2015} {This can be understood from the two contributions to the $dc$ voltage in ST-FMR in insulators: 1) spin pumping and 2) mixing of spin Hall magnetoresistance and the microwave signal. Both contributions follow a $\mathrm{cos}(\phi)\mathrm{sin}(2\phi)$ in-plane angular dependence, that is maximum at $\phi = 45^\circ$. \cite{Chiba_PRA_2014,Chiba_JAP_2015}} In order to excite spin dynamics by ST-FMR in the YIG nanowires, a $rf$ signal is passed through the Pt capping layer. Upon reaching the condition for ferromagnetic resonance, dynamics are driven resonantly, see Fig.~\ref{Fig_Setup}(c). Spin dynamics is governed by a modified Landau-Lifshitz-Gilbert equation \cite{Chiba_PRA_2014,Chiba_JAP_2015}
\begin{equation}
\label{mod_LLG}
\frac{d\mathbf{M}}{dt}= - \vert\gamma\vert \mathbf{M}\times \mathbf{H}_\mathrm{eff}+\alpha \mathbf{M}\times \frac{d\mathbf{M}}{dt}+\frac{\vert\gamma\vert \hbar}{2e M_\mathrm{s}d_\mathrm{F}}\mathbf{J}_\mathrm{s},
\end{equation}
where $\gamma$ 
is the gyromagnetic ratio, $\mathbf{H}_\mathrm{eff}=\mathbf{h}_\mathrm{rf}+\mathbf{H}_\mathrm{D}+\mathbf{H}_\mathrm{}$ is the effective magnetic field including the Oersted field $\mathbf{h}_\mathrm{rf}$, demagnetization fields $\mathbf{H}_\mathrm{D}$, and the externally applied magnetic field $\mathbf{H}$. $\alpha$ is the Gilbert damping parameter [the second term describes the damping torque $\tau_\alpha$, see Fig.~\ref{Fig_Setup}(b)] and $\mathbf{J}_\mathrm{s}$ is a transverse spin current at the interface generated by the SHE. $\mathbf{J}_\mathrm{s}$ has field-like, as well as anti-damping-like torque terms,\cite{Sklenar_PRB_2015} as illustrated in Fig.~\ref{Fig_Setup}(b). 

\section{Results and Discussion}

\begin{figure}
  \includegraphics[width=1\columnwidth]{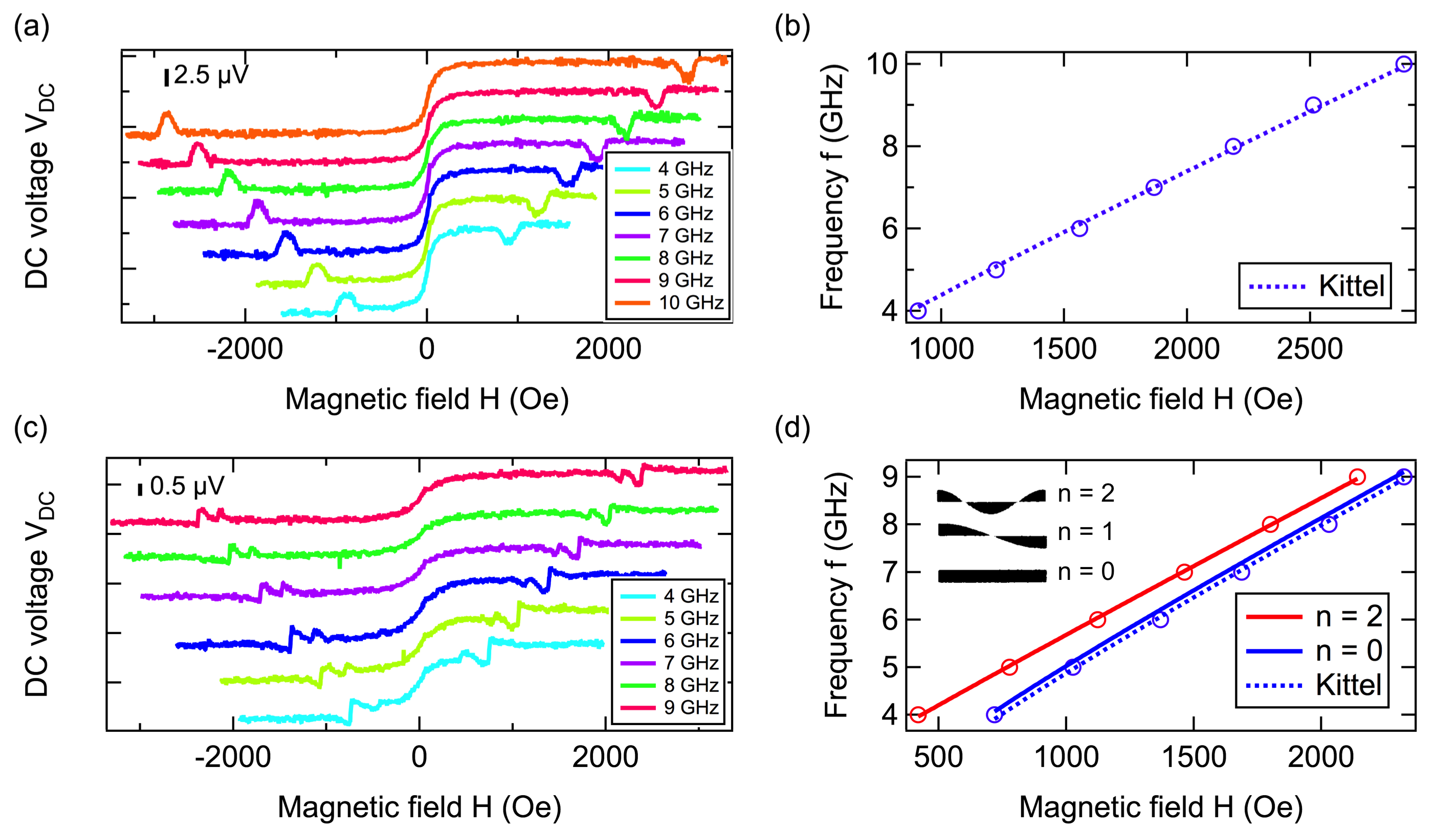}
  \caption{Room-temperature ST-FMR measurements. (a) Nanowire width: 765 nm. Voltage spectra at a fixed microwave power of 10 mW for different excitation frequencies. (b) Corresponding frequency-field relation that confirms the excitation of ferromagnetic resonance. {The dashed blue line shows a fit to the Kittel equation taking into account finite-size effects.} (c) Nanowire width: 365 nm. Spectra for a fixed microwave power of 1 mW. Clearly, a second mode due the lateral confinement appears for the thinner nanowires. (d) The energy of both modes increases with the applied magnetic field. {The solid lines are a fit to Eq.~(\ref{DE}) that considers quantization of spin-wave modes across the wire width. The inset shows the profile of the lateral standing modes. The result is compared to the Kittel equation with demagnetization (dashed).}}
  \label{Fig_Spec}
\end{figure}

Figure~\ref{Fig_Spec}(a,c) show the experimental results of ST-FMR on YIG/Pt nanowires [(a): 765 nm width, (c): 365 nm] at fixed microwave power [(a): 10 mW, (c): 1 mW] for different excitation frequencies. Clearly, upon realization of ferromagnetic resonance condition, a signal is observed. In case of the narrower wire width, we detect two modes. {In order to understand the underlying dynamics, we follow an approach that was originally presented in \cite{Jorzick_PRB_1999}. The appearance of the second mode in the spectrum of the 365 nm wire is due to a discretization of wavevectors across the wire width and we analyze the data in the following fashion: Since the long side of the nanowire and the applied magnetic field make an angle of 45$^\circ$, we consider an effective wire width which is perpendicular to the field (we use here 400 nm). Please note that micromagnetic simulations based on mumax3 \cite{mumax3} reveal that the magnetization is in good approximation aligned with the magnetic field in the here considered field range. The high-frequency mode is fitted to \cite{Jorzick_PRB_1999}
\begin{equation}
\label{DE}
f_\mathrm{DE}= \gamma/(2 \pi) [(H\cdot(H+4\pi M_\mathrm{S})+(2\pi M_\mathrm{S})^2\cdot(1-e^{-2 k t})]^{1/2},
\end{equation}
where $\gamma$ is the gyromagnetic ratio, $t$ is the YIG thickness, $M_\mathrm{S}$ is the saturation magnetization, $k =n\pi/ d$ is the quantized wavevector with $n = 0, 1, 2, ...$ and $d$ is the effective width. Equation~(\ref{DE}) is very similar to the result of the Damon-Eshbach formalism. A fit of Eq.~(\ref{DE}) to our data yields $M_\mathrm{S}$ = 176.77 emu/cm$^3$ and a quantization number n = 2 (2.05 $\pm$ 0.37), see red line in Fig.~\ref{Fig_Spec}(d). A fit of the low-frequency mode with $M_\mathrm{S}$ = 176.77 emu/cm$^3$ gives n = 0, see blue solid line in Fig.~\ref{Fig_Spec}(d). Although Eq.~(\ref{DE}) does not consider any finite-size effects a reasonably good agreement is found. As a comparison, we also show a fit to the Kittel equation which considers demagnetization fields in the same viewgraph. The blue dashed line is a fit to \cite{Ding_PRB_2011}:
\begin{equation}
\label{demag}
f= \gamma/(2 \pi) [(H+(1-2N_\mathrm{x})4\pi M_\mathrm{S})\cdot(H-N_\mathrm{x}4\pi M_\mathrm{S})]^{1/2},
\end{equation}
where $N_\mathrm{x}$ is the demagnetization factor in the direction of the applied field and $N_\mathrm{x}+N_\mathrm{y}=1$, with $N_\mathrm{y}$ -- demagnetization factor perpendicular to the field. Using $M_\mathrm{S}$ = 176.77 emu/cm$^3$, we find $N_\mathrm{x} = 0.02 \pm 3\times 10^{-3}$. As discussed in detail below, heating effects are significant in our wires and the data shown in Fig.~3(b) was collected at 10 mW, where a reduction of $M_\mathrm{S}$ cannot be neglected. In order to fit Fig.~\ref{Fig_Spec}(b), we assume that heating effects are negligible at the lowest microwave power (2.5 mW) we use in experiment and solve Eq.~(\ref{demag}) numerically with the previously found $M_\mathrm{S}$ = 176.77 emu/cm$^3$. We find $N_\mathrm{x} = 0.023$. The blue dashed line in Fig.~\ref{Fig_Spec}(b) shows a fit to Eq.~(\ref{demag}) using this value of $N_\mathrm{x}$. An effective magnetization of  $M_\mathrm{S}^\mathrm{eff}=129.65\pm2.88$ emu/cm$^3$ is found, which is reduced due to heating. Reduction of  $M_\mathrm{S}^\mathrm{eff}$ due to heating at high microwave powers will be discussed in detail below. By comparing the two wire widths it becomes evident that confinement effects play an important role for the narrower wires. }


 The offset visible in Fig.~\ref{Fig_Spec}(a,c) is attributed to the longitudinal spin-Seebeck effect caused by off-resonant heating {due to the $rf$ current flowing in the Pt layer}. \cite{Uchida_APL_2010,Rezende_PRB_2014,Miao_Adv_2016,
Jungfleisch_PRL_2016,Sklenar_PRB_2015} {The on-resonant and off-resonant signals have opposite polarities and we can therefore conclude that in case of the off-resonant signal a spin-Seebeck driven spin current flows from the Pt layer, which is heated up,  to the adjacent YIG layer.}

According to the model \cite{Chiba_JAP_2015,Chiba_PRA_2014}, two signals contribute to the $dc$ voltage: (1) spin pumping which manifests in a symmetric contribution to the Lorentzian lineshape; (2) spin Hall magnetoresistance which is a superimposed symmetric and antisymmetric Lorentzian curve. Figure~\ref{Fig_Setup}(c) shows the spectrum of the narrower wire width at 7 GHz and 1 mW. Clearly, the data can be fitted to a superimposed of symmetric and antisymmetric Lorentzian lineshape. This observation is not necessarily expected since the model assumes a macrospin model. It is surprising that it is possible to use generic fitting by Lorentzian curves even in the nanoscaled devices investigated here. We show that the ST-FMR concept can be applied to insulating nanomagnets and we limit ourselves to a qualitative description of the signal by Lorentzian lineshapes, rather than applying the exact model to extract material parameters. These studies can be found elsewhere. \cite{Jungfleisch_PRL_2016,Sklenar_PRB_2015} 
We note that the presence of both symmetric and antisymmetric components to the signal indicate that the dynamics is driven simultaneous by the Oersted field and spin-Hall spin torques. 

\begin{figure}
  \includegraphics[width=0.65\columnwidth]{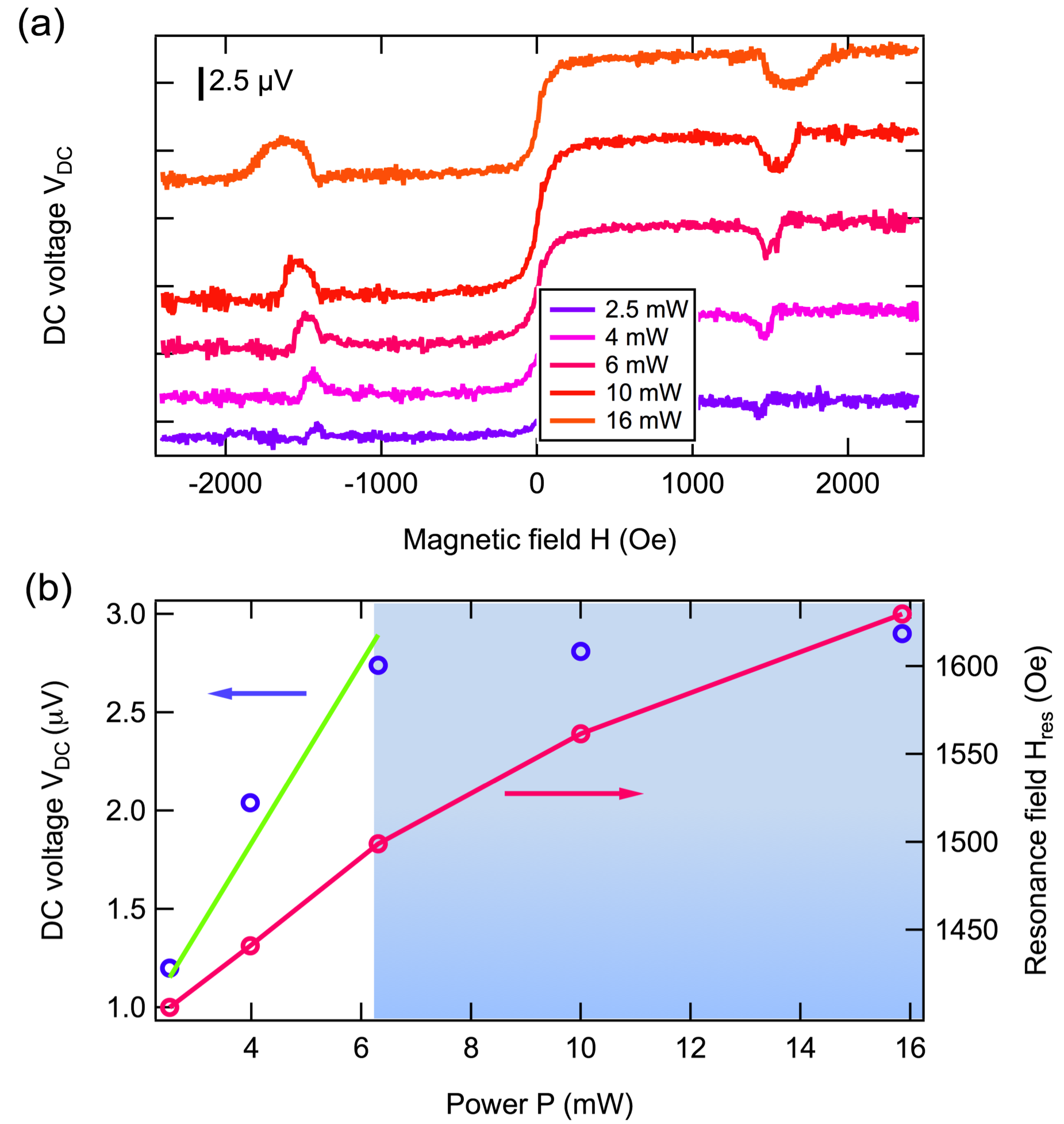}
  \caption{Power-dependent measurements, here nanowire width: 765 nm. (a) The amplitude of the $dc$ voltage increases with the applied power. A deformation of the peak shape is visible, which can be traced back to the onset of nonlinear effects. (b) Left: Peak maximum as a function of microwave power. In the low-power regime a linear behavior is observed (green line). At 6 mW the amplitude saturates (shaded area). Right: Microwave-power dependence of the observed ferromagnetic resonance field $H_\mathrm{res}$.}
  \label{Fig_Power}
\end{figure}

We also study the power dependence of the detected signal as is shown in Fig.~\ref{Fig_Power}. Figure.~\ref{Fig_Power}(a) illustrates the power-dependent spectra of the wider wire width of 765 nm at 6 GHz. The amplitude of the detected $dc$ voltage increases with the applied power. At the same time, a peak deformation occurs, which is common when nonlinear effects set in. \cite{Ando_PRL_2012,Lustikova_JAP_2015} To illustrate this more clearly we plot the peak amplitude and the resonance field as a function of the applied microwave power in Fig.~\ref{Fig_Power}(b). As is apparent from the viewgraph, the peak amplitude increases linearly with the applied power before it goes into saturation at approximately 
6 mW (shaded area). It is important to note that this \textit{threshold} power for the onset of nonlinear effects is much smaller than the threshold power usually observed in larger YIG samples. \cite{Ando_PRL_2012,Lustikova_JAP_2015,Sandweg_PRL_2011,Jungfleisch_PRL_2016,Jungfleisch_PRB_2015} However, as we will discuss in the following, strong heating effects are observed in our samples and it is difficult to unambiguously demonstrate that the saturation of the peak voltage is solely due to nonlinear phenomena in the spin system. Despite of whether the observed nonlinear behavior is due to nonlinear spin dynamics or due to a microwave-heating induced evolution of the detected $dc$ voltage, this nonlinearity could eventually be used in nanoscaled magnonic logic circuits based on insulators.

According to the ST-FMR theory, the detected ST-FMR voltage depends directly on the static magnetization component. This is also apparent from Fig.~\ref{Fig_Power}(b), right scale, where we plot the ferromagnetic resonance field. {Using Eq.~(\ref{demag}), we can estimate the change of $M_\mathrm{S}^\mathrm{eff}$ as function of the microwave power. A large reduction from 176.77 emu/cm$^3$ to 108.67 emu/cm$^3$ is found.
} It is well known that the effective magnetization changes at high driving powers because of three reasons \cite{Zhang_IEEE_1986,Zhang_JAP_1988,Lustikova_JAP_2015}: (1) Redistribution of energy to other spin-wave modes and spin-wave instability processes. (2) Opening of the precession cone angle. (3) Heating of the YIG layer due to microwave absorption. Different from conventional FMR measurements \cite{Zhang_IEEE_1986,Zhang_JAP_1988} and spin-pumping experiments \cite{Lustikova_JAP_2015,Sandweg_PRL_2011,Ando_PRL_2012}, we suspect that in our case the main reason for the change in the magnetization with power is strong heating effects. It is not likely that multi spin-wave processes or opening of the cone angle that are common reasons in FMR and spin-pumping experiments can cause such a large change in $4\pi M_\mathrm{eff}$ as observed here. To estimate the temperature rise in our sample, we assume that heating due to microwave absorption is the only reason for the change in the magnetization and this estimation will give us an upper limit for the temperature rise. 
Using Bloch's law $M(T) = M_0 (1-(T/T_\mathrm{C})^{3/2})$ we estimate the temperature rise in the YIG nanowire to be {$\Delta T \approx 105$~K} for $P=$ 16 mW. This value is much larger than the temperature rise found in FMR measurements of micrometer-thick YIG slabs and spin-pumping experiments. \cite{Zhang_IEEE_1986,Lustikova_JAP_2015} This observation is reasonable since the microwave signal in our experiment is passed through the Pt layer that is in direct contact to the YIG nanowire, whereas in FMR and spin-pumping experiments an antenna or cavity is used to excite the spin dynamics. {We do not only observe unusual behavior of the resonant signal, but also of the off-resonant voltage caused by the longitudinal SSE visible as an offset in Fig.~\ref{Fig_Power}(a): Usually, the SSE signal increases with the applied power/heat. However, in our case we observe a saturation, followed by a reduction of the signal. This observation might be a further indication of a change in the magnetization at high power due to heating.}

\begin{figure}
  \includegraphics[width=0.75\columnwidth]{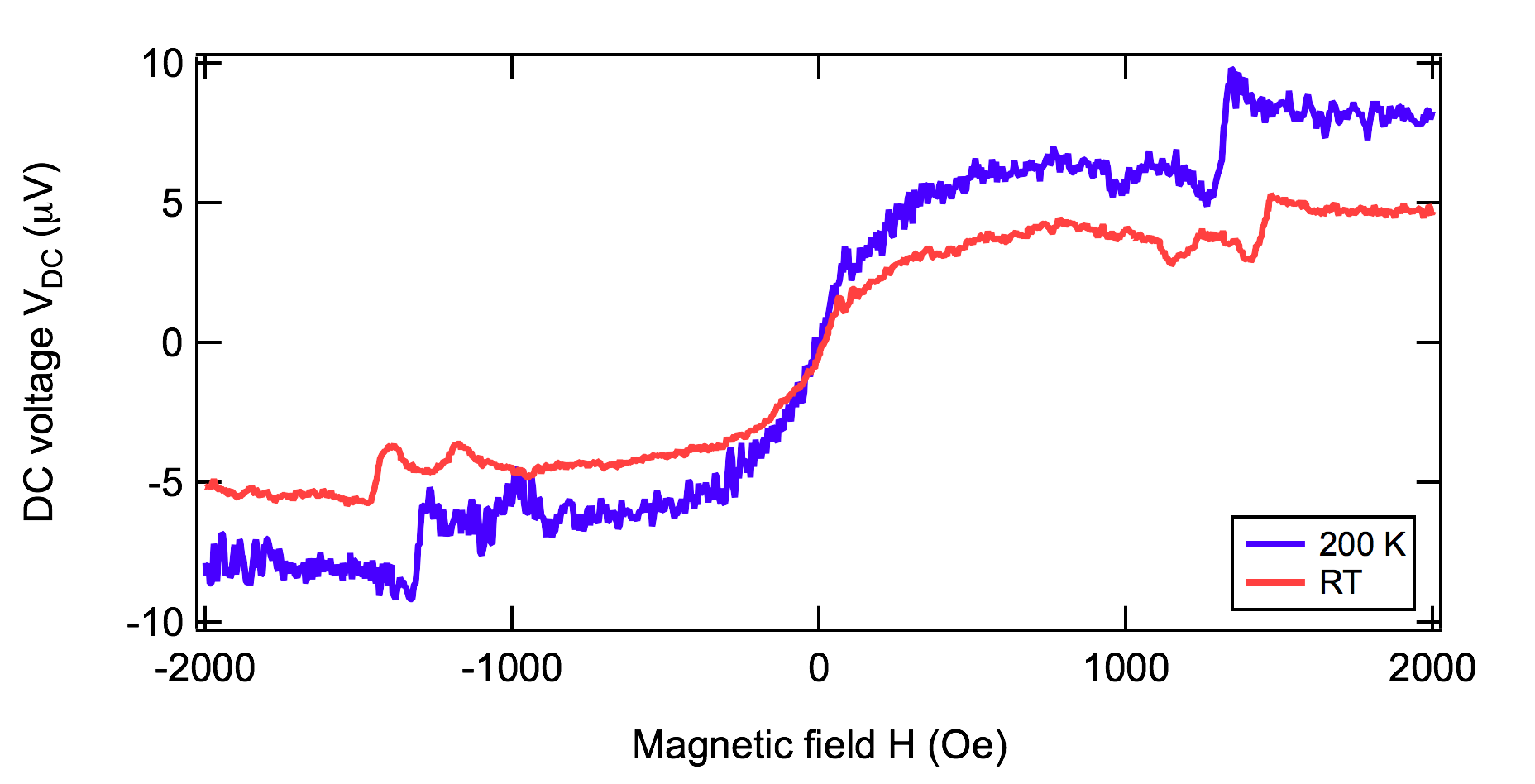}
  \caption{Temperature dependence of $dc$ voltage spectra for the 365 nm wire. Excitation frequency: 6 GHz, applied power: 4 mW.}
  \label{Fig_Temp}
\end{figure}

We further characterized our samples at low temperatures. Figure~\ref{Fig_Temp} shows the ST-FMR spectra of the 365-nm nanowire at room temperature (RT) and 200 K. As is apparent from the figure, the signal persists at 200 K. Although the modes contributing to the signal do not change, the amplitude, offset voltage and resonance field differ. The amplitude at low temperature increases indicating either a better microwave transmission of the ST-FMR circuitry or a reduced magnetic damping due to less thermally activated magnons. The increase of the offset voltage due to the longitudinal SSE is likely due to a larger temperature gradient across the sample, which develops at low surrounding temperatures. This can be understood as follows: At 200 K the GGG substrate and the YIG wires are at a smaller temperature than at RT, whereas the temperature of the Pt layer is almost the same as at RT since the microwave signal is passed through it. \cite{Uchida_APL_2010} Thus, the temperature gradient is increased leading to larger spin current emission from the Pt into the YIG layer. We also note that the resonance field at 200 K shifts to lower fields, which is due to an increase of the effective magnetization and can be considered as an indirect confirmation of our explanation of Fig.~\ref{Fig_Power}(b) discussed in the previous paragraph.

\section{Conclusion}
In conclusion, we demonstrated ST-FMR in nanoscaled YIG/Pt devices. A rectified $dc$ voltage was measured under resonance condition by a SMR-mediated ST-FMR mechanism. Despite the fact that theoretical works assume a macrospin model to explain the involved effects, we showed here that the phenomenological explanation of driving the dynamics by field-like and anti-damping-like torque components can be applied to nanoscaled devices. Down-scaling of sample dimensions further reveals the importance of lateral confinements of the magnetic material to the dynamic response. 
A large shift in the resonance field is observed at high microwave powers, which is explained by a lowering of the magnetization due to microwave absorption. The estimated temperature rise of the 765-nm nanowire is roughly 105 K.
Comparison of $dc$-voltage spectra at room temperature and at 200 K reveals an increase of the magnitude of the detected output, which is either due to a lower magnetic damping (i.e., less thermally activated magnons) or better microwave transmission properties of the ST-FMR circuitry (i.e., electric susceptibility of the GGG substrate). {The realization of ST-FMR in YIG/Pt nanowires is one step closer to the implementation of integrated magnonic logic devices based on insulators since it combines spin dynamics in low-damping nanomagnets with spintronics effects occurring at the Pt interface.}
The presented results are the first stepping stones toward the realization of magnon spintronics in nanoscaled devices based on magnetic insulators.

\begin{acknowledgement}


This work was supported by the U.S. Department of Energy, Office of Science, Materials Science and Engineering Division. Lithography was carried out at the Center for Nanoscale Materials, an Office of Science user facility, which is supported by DOE, Office of Science, Basic Energy Science under Contract No. DE-AC02-06CH11357.

\end{acknowledgement}

%
%
%

\bibliography{Nano_Letters}

\end{document}